\documentclass{article}
\usepackage{axodraw,amsmath,alltt,feynarts,pifont,multicol,a4wide}

\newcommand{\FA}{\textsl{FeynArts}}
\newcommand{\mma}{\textsl{Mathematica}}
\newcommand{\lbrac}{\symbol{123}}
\newcommand{\rbrac}{\symbol{125}}

\newcommand{\eg}{e.g.\ }
\newcommand{\ie}{i.e.\ }

\renewcommand{\FADiagramLabelSize}{\scriptsize}

\makeatletter

\def\reportno#1{\gdef\@reportno{#1}}

\def\@maketitle{%
  \hfill{\small\begin{tabular}[t]{r}%
    \@reportno
  \end{tabular}\par}%
  \vskip 2em%
  \begin{center}%
    \let\footnote\thanks%
    {\large\@title\par}%
    \vskip 1.5em%
    \lineskip .5em%
    \begin{tabular}[t]{c}%
      \@author
    \end{tabular}\par%
    \vskip 1em%
    \@date%
  \end{center}%
  \par
  \vskip 1.5em}

\makeatother

\begin{document}

\title{Generating Feynman Diagrams and Amplitudes with \FA\ 3}

\author{Thomas Hahn \\
	{\it Institut f\"ur Theoretische Physik} \\
	{\it Universit\"at Karlsruhe} \\
	{\it D--76128 Karlsruhe, Germany}}

\reportno{KA--TP--23--2000\\hep-ph/0012260}

\date{December 20, 2000}

\maketitle

\begin{abstract}
This paper describes the \mma\ package \FA\ used for the generation and
visualization of Feynman diagrams and amplitudes. The main features of
version 3 are: generation of diagrams in three levels, user-definable
model files, support for supersymmetric models, and publication-quality
Feynman diagrams in PostScript or \LaTeX.\\[1ex]
\end{abstract}


\section*{PROGRAM SUMMARY}

\newdimen\oldparindent%
\oldparindent=\parindent%
\parindent=0sp%

\begin{multicols}{2}

{\it Title of program:}
\FA

\medskip

{\it Catalogue identifier:}

\medskip

{\it Program obtainable from:}
CPC Program Library, Queen's University of Belfast, N. Ireland, and
{\tt http://www.feynarts.de}

\medskip

{\it Computer for which the program is designed and others on which is has
been tested:}\\
{\it Designed for:}
platforms on which \mma\ and Java are available \\
{\it Tested on:}
Intel-based PCs, DEC Alpha workstations

\medskip

{\it Operating systems or monitors under which the program has been
tested:}
Linux, Tru64 Unix

\medskip

{\it Programming language used:}
\mma, Java

\medskip

{\it Memory required to execute with typical data:}
8M words

\medskip

{\it No.\ of bits in a word:}
8

\medskip

{\it No.\ of processors used:}
1

\medskip

{\it Has the code been vectorized or parallelized?}
No

\medskip

{\it No.\ of bytes in distributed program, including test data, etc.:}
$\sim$ 400 K bytes

\medskip

{\it Distribution format:}
gzipped tar archive

\medskip

{\it Keywords:}
Feynman diagrams, Perturbation theory, Quantum field theory,
Green's functions, $S$-matrix elements.

\medskip

{\it Nature of the physical problem:}
Feynman-diagrammatic computations in field theory.

\medskip

{\it Method of solution:}
\FA\ works in three steps: 1) creation of the topologies, 2) insertion of
fields into the topologies, 3) application of the Feynman rules to produce
Feynman amplitudes. Information about the physical model, such as the
Feynman rules, is provided in a so-called model file.

\medskip

{\it Typical running time:}
About a minute to generate all amplitudes for a one-loop, $2\to 2$ process 
in the electroweak Standard Model.

\medskip

{\it Unusual features of the program:}
\FA\ can produce high-quality images of the Feynman diagrams \eg in
PostScript or \LaTeX\ format for inclusion in publications.

\medskip

{\it Restrictions on the complexity:}
Currently diagrams up to three loops can be generated. Model files other
than the Standard Model and QCD (and soon also the Minimal Supersymmetric
Standard Model) are not contained in \FA\ and must be set up by the
user.

\end{multicols}

\parindent=\oldparindent%


\section*{LONG WRITE-UP}

\section{Introduction}

Much as field theorists would love to abandon them in favour of a less
laborious technique, Feynman diagrams \cite{Ve94} are unlikely to become
extinct in the forseeable future. Adding to the amount of work,
increasingly precise experimental data nowadays mandate calculations
involving substantial numbers of Feynman diagrams. Beginning in the 1960s,
people started following up on the rather obvious idea of letting the
computer do all those involved calculations, and indeed one of the first
computer-algebra systems, {\sc Schoonschip}, was invented for precisely
this purpose by Nobel-laureate Martinus Veltman.

This paper describes another part of this quest: the \mma\ package \FA,
used for the generation and visualization of Feynman diagrams and
amplitudes. It performs the first step of a field-theoretic perturbative
calculation, leaving simplification and numerical evaluation of the
amplitudes to other programs, \eg \cite{MeBD91, HaP98}. Programs with a
similar functionality are QGRAF \cite{No93}, the grc part of GRACE
\cite{Yu00}, CompHEP \cite{BoDIPS94}, and to a lesser extent MadGraph
\cite{StL94}.

A bit of history: \FA\ started out in 1990 as a Macsyma code written by
Eck and K\"ublbeck which could produce diagrams in the Standard Model
\cite{KuBD90}, but it soon was ported to the \mma\ platform. In 1995, Eck
and K\"ublbeck designed the second version to be a fully general diagram
generator. To achieve this, they implemented some decisive new ideas
\cite{Eck95}, the most important one being the generation of diagrams in
three levels. The program was taken up again in 1998 by Hahn who developed
version 2.2. The well-designed conceptual framework was kept, but the
actual code was reprogrammed almost entirely to make it more efficient and
a user-friendly topology editor was added. The current third version
features in particular significantly improved graphics. For example, it is
now very easy to include Feynman diagrams produced by \FA\ in a \LaTeX\
document.

\medskip

The main features of \FA\ are:
\begin{itemize}
\item
The generation of diagrams is possible in three levels: generic fields,
classes of fields, or specific particles.

\item
The model information is contained in two special files:
The {\em generic model file} defines the representation of the
kinematical quantities like spinors or vector fields. The {\em classes 
model file} sets up the particle content and specifies the actual
couplings. Since the user can create own model files, the applicability
of \FA\ is virtually unlimited within perturbative quantum field
theory.

\item
In addition to ordinary diagrams, \FA\ can generate counter-term diagrams
and diagrams with placeholders for one-particle irreducible vertex
functions (skeleton diagrams).

\item
\FA\ employs the so-called ``flipping-rule'' algorithm \cite{DeEHK92} to
concatenate fermion chains. This algorithm is unique in that it works
also for Majorana fermions and for the fermion-number-violating couplings
they entail and hence allows supersymmetric models to be implemented.

\item
Restrictions of the type ``field $X$ is not allowed in loops'' can be
applied. This is necessary \eg for the background-field formulation of a
field theory.

\item
Vertices of arbitrary adjacency\footnote{The adjacency of a vertex is
the number of lines that join at the vertex.}, required for effective
theories, are allowed.

\item
Mixing propagators, such as appear in non-$R_\xi$-gauges, are supported.

\item
\FA\ produces publication-quality Feynman diagrams in PostScript or
\LaTeX\ in a format that allows easy customization.
\end{itemize}
These features have been introduced in version 2 and some of them have
received considerable improvements in version 3. The user interface, on
the other hand, has through the versions suffered only minor and mostly
backward-compatible changes, and the major functions can still be used in
essentially the same way as in version 1.

This paper is divided into two parts: Sect.\ \ref{sect:usage} gives a
brief survey of the main functions from a user's perspective. The concepts
of the computer-algebraic generation of Feynman diagrams and amplitudes
and their implementation in \FA\ are discussed in Sect.\
\ref{sect:concepts}.


\section{Using \FA}
\label{sect:usage}

\begin{figure}[t]
\begin{center}
\begin{small}
\unitlength=1bp%
\begin{picture}(190,275)(35,16)
\SetScale{.6}
\SetWidth{1.5}
\ArrowLine(150,411)(150,380)
\ArrowLine(150,335)(150,308)
\ArrowLine(150,251)(150,222)
\ArrowLine(150,175)(150,148)
\ArrowLine(150,101)(150,72)

\Line(200,50)(280,50)
\ArrowLine(279,50)(280,50)
\Text(180,32.8)[lb]{further}
\Text(180,28.8)[lt]{processing}

\ArrowArcn(210,260)(100,90,27)
\ArrowArc(210,300)(100,-90,-27)
\SetWidth{.5}

\GBox(62,410)(238,488){.9}
\Text(90,280.8)[b]{Find all distinct ways}
\Text(90,270.4)[b]{of connecting incoming}
\Text(90,260)[b]{and outgoing lines}
\Text(90,257.6)[t]{({\tt CreateTopologies})}
\GOval(150,360)(25,65)(0){1}
\Text(90,215.2)[]{Topologies}
\GBox(65,250)(235,310){.9}
\Text(90,174.4)[b]{Determine all allowed\vphantom{p}}
\Text(90,164)[b]{combinations of fields\vphantom{p}}
\Text(90,161.6)[t]{({\tt InsertFields})}
\GBox(248,250)(377,310){.9}
\Text(187,169.6)[b]{Draw the results\vphantom{p}}
\Text(187,165.6)[t]{({\tt Paint})}
\GOval(150,200)(25,65)(0){1}
\Text(90,119.2)[]{Diagrams}
\GBox(55,100)(245,150){.9}
\Text(90,76.8)[b]{Apply the Feynman rules}
\Text(90,72.8)[t]{({\tt CreateFeynAmp})}
\GOval(150,50)(25,65)(0){1}
\Text(90,29.6)[]{Amplitudes}
\end{picture}
\end{small}
\end{center}
\vskip -15bp%
\caption{\label{fig:feynarts}%
Flowchart for the generation of Feynman amplitudes with \FA.}
\end{figure}
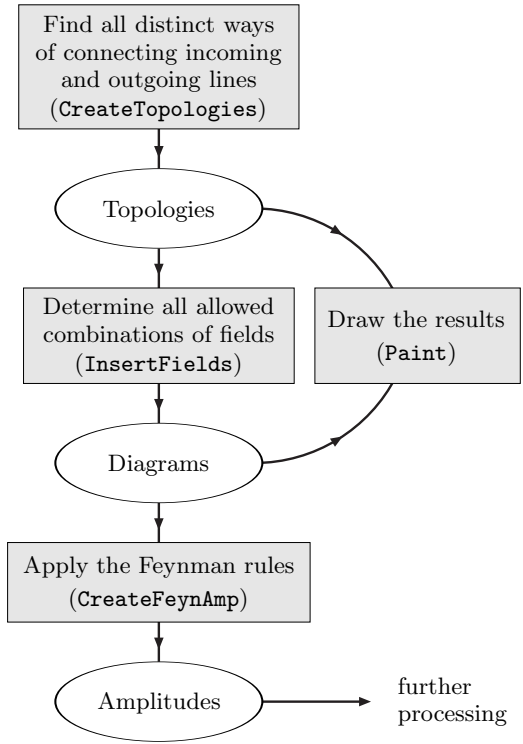

\FA\ works in the three basic steps sketched in Fig.~\ref{fig:feynarts}.

The first step is to create all different topologies for a given number of
loops and external legs. The following call to {\tt CreateTopologies}
creates for example all one-particle-irreducible (1PI) one-loop topologies
for a $1\to 2$ process. This is done by generating all one-loop $1\to 2$
topologies and then excluding the reducible ones:
\begin{quote}
\begin{verbatim}
top = CreateTopologies[1, 1 -> 2, ExcludeTopologies -> Internal]
\end{verbatim}
\end{quote}
The output of {\tt CreateTopologies} is an internal data structure called
a {\tt TopologyList}. As an example, the first topology in the
{\tt TopologyList} just created is shown here:
\begin{quote}
\begin{verbatim}
Topology[1][ Propagator[Incoming][Vertex[1][1], Vertex[3][4]], 
             Propagator[Outgoing][Vertex[1][2], Vertex[3][5]], 
             Propagator[Outgoing][Vertex[1][3], Vertex[3][6]], 
             Propagator[Loop[1]][Vertex[3][4], Vertex[3][5]], 
             Propagator[Loop[1]][Vertex[3][4], Vertex[3][6]], 
             Propagator[Loop[1]][Vertex[3][5], Vertex[3][6]] ]
\end{verbatim}
\end{quote}
A much nicer way to visualize the {\tt TopologyList} in {\tt top} is to
paint it with {\tt Paint[top]}.
\begin{quote}
\begin{scriptsize}
\begin{feynartspicture}(288,336)(4,4.3)
\FALabel(44.,91.96)[]{\large $1\quad \to\quad 2$}

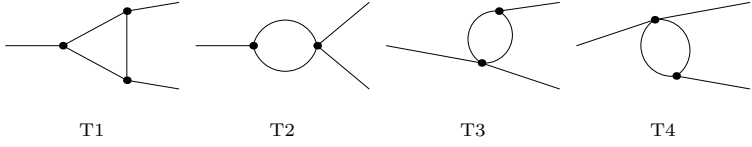
\FADiagram{T1}
\FAProp(0.,10.)(6.6,10.)(0.,){/Straight}{0}
\FAProp(20.,15.)(14.,14.)(0.,){/Straight}{0}
\FAProp(20.,5.)(14.,6.)(0.,){/Straight}{0}
\FAProp(6.6,10.)(14.,14.)(0.,){/Straight}{0}
\FAProp(6.6,10.)(14.,6.)(0.,){/Straight}{0}
\FAProp(14.,14.)(14.,6.)(0.,){/Straight}{0}
\FAVert(6.6,10.){0}
\FAVert(14.,14.){0}
\FAVert(14.,6.){0}

\FADiagram{T2}
\FAProp(0.,10.)(6.6,10.)(0.,){/Straight}{0}
\FAProp(20.,15.)(14.,10.)(0.,){/Straight}{0}
\FAProp(20.,5.)(14.,10.)(0.,){/Straight}{0}
\FAProp(6.6,10.)(14.,10.)(0.8,){/Straight}{0}
\FAProp(6.6,10.)(14.,10.)(-0.8,){/Straight}{0}
\FAVert(6.6,10.){0}
\FAVert(14.,10.){0}

\FADiagram{T3}
\FAProp(0.,10.)(11.,8.)(0.,){/Straight}{0}
\FAProp(20.,15.)(13.,14.)(0.,){/Straight}{0}
\FAProp(20.,5.)(11.,8.)(0.,){/Straight}{0}
\FAProp(13.,14.)(11.,8.)(0.8,){/Straight}{0}
\FAProp(13.,14.)(11.,8.)(-0.8,){/Straight}{0}
\FAVert(13.,14.){0}
\FAVert(11.,8.){0}

\FADiagram{T4}
\FAProp(0.,10.)(9.,13.)(0.,){/Straight}{0}
\FAProp(20.,15.)(9.,13.)(0.,){/Straight}{0}
\FAProp(20.,5.)(11.5,6.5)(0.,){/Straight}{0}
\FAProp(11.5,6.5)(9.,13.)(0.8,){/Straight}{0}
\FAProp(11.5,6.5)(9.,13.)(-0.8,){/Straight}{0}
\FAVert(11.5,6.5){0}
\FAVert(9.,13.){0}
\end{feynartspicture}
\end{scriptsize}
\vspace*{-7.5cm}
\end{quote}

In the second step, the actual particles in the model have to be
distributed over the topologies in all allowed ways. For example, the
diagrams for $Z\to b\bar b$ are produced with
\begin{quote}
\begin{verbatim}
ins = InsertFields[top, V[2] -> {F[4,{3}], -F[4,{3}]},
        Model -> SM, InsertionLevel -> {Classes}]
\end{verbatim}
\end{quote}
where {\tt F[4,\,\lbrac 3\rbrac]} is the $b$-quark,
{\tt -F[4,\,\lbrac 3\rbrac]} its antiparticle, and \verb=V[2]= the $Z$
boson. The model information is taken from the file {\tt SM.mod}.
The insertion level tells {\tt InsertFields} how detailed each inserted
field should be specified: {\tt \lbrac Classes\rbrac} means that classes
of fields should not be expanded. In the Standard Model, for example, the
fermions are arranged in classes, so there will be only one diagram for
\eg an up-type quark $u_i$ instead of three for its members $u_1 = u$,
$u_2 = c$, $u_3 = t$ (see Sect.\ \ref{sect:levels} for more information on
field levels).

The output of {\tt InsertFields} is again a {\tt TopologyList} which is
now supplemented with the field information. Its printed form is lengthy
and looks rather unappealing, but it can be drawn with {\tt Paint[ins]}
which results in
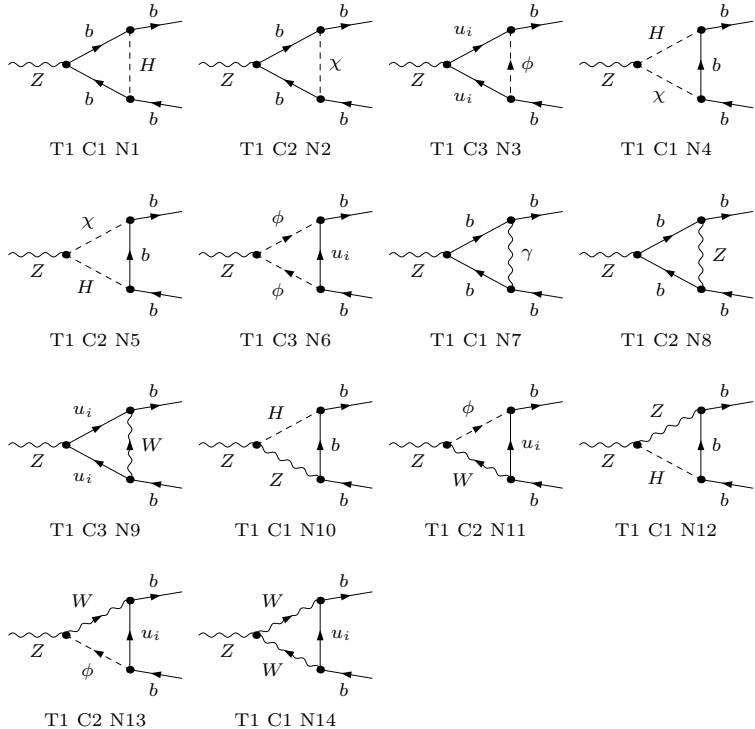
\begin{quote}
\begin{scriptsize}
\begin{feynartspicture}(288,336)(4,4.3)
\FALabel(44.,91.96)[]{\large $Z\quad \to\quad b\quad b$}

\FADiagram{T1 C1 N1}
\FAProp(0.,10.)(6.6,10.)(0.,){/Sine}{0}
\FALabel(3.3,8.93)[t]{$Z$}
\FAProp(20.,15.)(14.,14.)(0.,){/Straight}{-1}
\FALabel(16.7452,15.5489)[b]{$b$}
\FAProp(20.,5.)(14.,6.)(0.,){/Straight}{1}
\FALabel(16.7452,4.45109)[t]{$b$}
\FAProp(6.6,10.)(14.,14.)(0.,){/Straight}{1}
\FALabel(9.85274,13.2354)[br]{$b$}
\FAProp(6.6,10.)(14.,6.)(0.,){/Straight}{-1}
\FALabel(9.85274,6.76457)[tr]{$b$}
\FAProp(14.,14.)(14.,6.)(0.,){/ScalarDash}{0}
\FALabel(15.024,10.)[l]{$H$}
\FAVert(6.6,10.){0}
\FAVert(14.,14.){0}
\FAVert(14.,6.){0}

\FADiagram{T1 C2 N2}
\FAProp(0.,10.)(6.6,10.)(0.,){/Sine}{0}
\FALabel(3.3,8.93)[t]{$Z$}
\FAProp(20.,15.)(14.,14.)(0.,){/Straight}{-1}
\FALabel(16.7452,15.5489)[b]{$b$}
\FAProp(20.,5.)(14.,6.)(0.,){/Straight}{1}
\FALabel(16.7452,4.45109)[t]{$b$}
\FAProp(6.6,10.)(14.,14.)(0.,){/Straight}{1}
\FALabel(9.85274,13.2354)[br]{$b$}
\FAProp(6.6,10.)(14.,6.)(0.,){/Straight}{-1}
\FALabel(9.85274,6.76457)[tr]{$b$}
\FAProp(14.,14.)(14.,6.)(0.,){/ScalarDash}{0}
\FALabel(15.024,10.)[l]{$\chi$}
\FAVert(6.6,10.){0}
\FAVert(14.,14.){0}
\FAVert(14.,6.){0}

\FADiagram{T1 C3 N3}
\FAProp(0.,10.)(6.6,10.)(0.,){/Sine}{0}
\FALabel(3.3,8.93)[t]{$Z$}
\FAProp(20.,15.)(14.,14.)(0.,){/Straight}{-1}
\FALabel(16.7452,15.5489)[b]{$b$}
\FAProp(20.,5.)(14.,6.)(0.,){/Straight}{1}
\FALabel(16.7452,4.45109)[t]{$b$}
\FAProp(6.6,10.)(14.,14.)(0.,){/Straight}{1}
\FALabel(9.85274,13.2354)[br]{$u_i$}
\FAProp(6.6,10.)(14.,6.)(0.,){/Straight}{-1}
\FALabel(9.85274,6.76457)[tr]{$u_i$}
\FAProp(14.,14.)(14.,6.)(0.,){/ScalarDash}{-1}
\FALabel(15.274,10.)[l]{$\phi$}
\FAVert(6.6,10.){0}
\FAVert(14.,14.){0}
\FAVert(14.,6.){0}

\FADiagram{T1 C1 N4}
\FAProp(0.,10.)(6.6,10.)(0.,){/Sine}{0}
\FALabel(3.3,8.93)[t]{$Z$}
\FAProp(20.,15.)(14.,14.)(0.,){/Straight}{-1}
\FALabel(16.7452,15.5489)[b]{$b$}
\FAProp(20.,5.)(14.,6.)(0.,){/Straight}{1}
\FALabel(16.7452,4.45109)[t]{$b$}
\FAProp(6.6,10.)(14.,14.)(0.,){/ScalarDash}{0}
\FALabel(9.97162,13.0155)[br]{$H$}
\FAProp(6.6,10.)(14.,6.)(0.,){/ScalarDash}{0}
\FALabel(9.97162,6.9845)[tr]{$\chi$}
\FAProp(14.,14.)(14.,6.)(0.,){/Straight}{-1}
\FALabel(15.274,10.)[l]{$b$}
\FAVert(6.6,10.){0}
\FAVert(14.,14.){0}
\FAVert(14.,6.){0}

\FADiagram{T1 C2 N5}
\FAProp(0.,10.)(6.6,10.)(0.,){/Sine}{0}
\FALabel(3.3,8.93)[t]{$Z$}
\FAProp(20.,15.)(14.,14.)(0.,){/Straight}{-1}
\FALabel(16.7452,15.5489)[b]{$b$}
\FAProp(20.,5.)(14.,6.)(0.,){/Straight}{1}
\FALabel(16.7452,4.45109)[t]{$b$}
\FAProp(6.6,10.)(14.,14.)(0.,){/ScalarDash}{0}
\FALabel(9.97162,13.0155)[br]{$\chi$}
\FAProp(6.6,10.)(14.,6.)(0.,){/ScalarDash}{0}
\FALabel(9.97162,6.9845)[tr]{$H$}
\FAProp(14.,14.)(14.,6.)(0.,){/Straight}{-1}
\FALabel(15.274,10.)[l]{$b$}
\FAVert(6.6,10.){0}
\FAVert(14.,14.){0}
\FAVert(14.,6.){0}

\FADiagram{T1 C3 N6}
\FAProp(0.,10.)(6.6,10.)(0.,){/Sine}{0}
\FALabel(3.3,8.93)[t]{$Z$}
\FAProp(20.,15.)(14.,14.)(0.,){/Straight}{-1}
\FALabel(16.7452,15.5489)[b]{$b$}
\FAProp(20.,5.)(14.,6.)(0.,){/Straight}{1}
\FALabel(16.7452,4.45109)[t]{$b$}
\FAProp(6.6,10.)(14.,14.)(0.,){/ScalarDash}{1}
\FALabel(9.85274,13.2354)[br]{$\phi$}
\FAProp(6.6,10.)(14.,6.)(0.,){/ScalarDash}{-1}
\FALabel(9.85274,6.76457)[tr]{$\phi$}
\FAProp(14.,14.)(14.,6.)(0.,){/Straight}{-1}
\FALabel(15.274,10.)[l]{$u_i$}
\FAVert(6.6,10.){0}
\FAVert(14.,14.){0}
\FAVert(14.,6.){0}

\FADiagram{T1 C1 N7}
\FAProp(0.,10.)(6.6,10.)(0.,){/Sine}{0}
\FALabel(3.3,8.93)[t]{$Z$}
\FAProp(20.,15.)(14.,14.)(0.,){/Straight}{-1}
\FALabel(16.7452,15.5489)[b]{$b$}
\FAProp(20.,5.)(14.,6.)(0.,){/Straight}{1}
\FALabel(16.7452,4.45109)[t]{$b$}
\FAProp(6.6,10.)(14.,14.)(0.,){/Straight}{1}
\FALabel(9.85274,13.2354)[br]{$b$}
\FAProp(6.6,10.)(14.,6.)(0.,){/Straight}{-1}
\FALabel(9.85274,6.76457)[tr]{$b$}
\FAProp(14.,14.)(14.,6.)(0.,){/Sine}{0}
\FALabel(15.274,10.)[l]{$\gamma$}
\FAVert(6.6,10.){0}
\FAVert(14.,14.){0}
\FAVert(14.,6.){0}

\FADiagram{T1 C2 N8}
\FAProp(0.,10.)(6.6,10.)(0.,){/Sine}{0}
\FALabel(3.3,8.93)[t]{$Z$}
\FAProp(20.,15.)(14.,14.)(0.,){/Straight}{-1}
\FALabel(16.7452,15.5489)[b]{$b$}
\FAProp(20.,5.)(14.,6.)(0.,){/Straight}{1}
\FALabel(16.7452,4.45109)[t]{$b$}
\FAProp(6.6,10.)(14.,14.)(0.,){/Straight}{1}
\FALabel(9.85274,13.2354)[br]{$b$}
\FAProp(6.6,10.)(14.,6.)(0.,){/Straight}{-1}
\FALabel(9.85274,6.76457)[tr]{$b$}
\FAProp(14.,14.)(14.,6.)(0.,){/Sine}{0}
\FALabel(15.274,10.)[l]{$Z$}
\FAVert(6.6,10.){0}
\FAVert(14.,14.){0}
\FAVert(14.,6.){0}

\FADiagram{T1 C3 N9}
\FAProp(0.,10.)(6.6,10.)(0.,){/Sine}{0}
\FALabel(3.3,8.93)[t]{$Z$}
\FAProp(20.,15.)(14.,14.)(0.,){/Straight}{-1}
\FALabel(16.7452,15.5489)[b]{$b$}
\FAProp(20.,5.)(14.,6.)(0.,){/Straight}{1}
\FALabel(16.7452,4.45109)[t]{$b$}
\FAProp(6.6,10.)(14.,14.)(0.,){/Straight}{1}
\FALabel(9.85274,13.2354)[br]{$u_i$}
\FAProp(6.6,10.)(14.,6.)(0.,){/Straight}{-1}
\FALabel(9.85274,6.76457)[tr]{$u_i$}
\FAProp(14.,14.)(14.,6.)(0.,){/Sine}{-1}
\FALabel(15.274,10.)[l]{$W$}
\FAVert(6.6,10.){0}
\FAVert(14.,14.){0}
\FAVert(14.,6.){0}

\FADiagram{T1 C1 N10}
\FAProp(0.,10.)(6.6,10.)(0.,){/Sine}{0}
\FALabel(3.3,8.93)[t]{$Z$}
\FAProp(20.,15.)(14.,14.)(0.,){/Straight}{-1}
\FALabel(16.7452,15.5489)[b]{$b$}
\FAProp(20.,5.)(14.,6.)(0.,){/Straight}{1}
\FALabel(16.7452,4.45109)[t]{$b$}
\FAProp(6.6,10.)(14.,14.)(0.,){/ScalarDash}{0}
\FALabel(9.97162,13.0155)[br]{$H$}
\FAProp(6.6,10.)(14.,6.)(0.,){/Sine}{0}
\FALabel(9.85274,6.76457)[tr]{$Z$}
\FAProp(14.,14.)(14.,6.)(0.,){/Straight}{-1}
\FALabel(15.274,10.)[l]{$b$}
\FAVert(6.6,10.){0}
\FAVert(14.,14.){0}
\FAVert(14.,6.){0}

\FADiagram{T1 C2 N11}
\FAProp(0.,10.)(6.6,10.)(0.,){/Sine}{0}
\FALabel(3.3,8.93)[t]{$Z$}
\FAProp(20.,15.)(14.,14.)(0.,){/Straight}{-1}
\FALabel(16.7452,15.5489)[b]{$b$}
\FAProp(20.,5.)(14.,6.)(0.,){/Straight}{1}
\FALabel(16.7452,4.45109)[t]{$b$}
\FAProp(6.6,10.)(14.,14.)(0.,){/ScalarDash}{1}
\FALabel(9.85274,13.2354)[br]{$\phi$}
\FAProp(6.6,10.)(14.,6.)(0.,){/Sine}{-1}
\FALabel(9.85274,6.76457)[tr]{$W$}
\FAProp(14.,14.)(14.,6.)(0.,){/Straight}{-1}
\FALabel(15.274,10.)[l]{$u_i$}
\FAVert(6.6,10.){0}
\FAVert(14.,14.){0}
\FAVert(14.,6.){0}

\FADiagram{T1 C1 N12}
\FAProp(0.,10.)(6.6,10.)(0.,){/Sine}{0}
\FALabel(3.3,8.93)[t]{$Z$}
\FAProp(20.,15.)(14.,14.)(0.,){/Straight}{-1}
\FALabel(16.7452,15.5489)[b]{$b$}
\FAProp(20.,5.)(14.,6.)(0.,){/Straight}{1}
\FALabel(16.7452,4.45109)[t]{$b$}
\FAProp(6.6,10.)(14.,14.)(0.,){/Sine}{0}
\FALabel(9.85274,13.2354)[br]{$Z$}
\FAProp(6.6,10.)(14.,6.)(0.,){/ScalarDash}{0}
\FALabel(9.97162,6.9845)[tr]{$H$}
\FAProp(14.,14.)(14.,6.)(0.,){/Straight}{-1}
\FALabel(15.274,10.)[l]{$b$}
\FAVert(6.6,10.){0}
\FAVert(14.,14.){0}
\FAVert(14.,6.){0}

\FADiagram{T1 C2 N13}
\FAProp(0.,10.)(6.6,10.)(0.,){/Sine}{0}
\FALabel(3.3,8.93)[t]{$Z$}
\FAProp(20.,15.)(14.,14.)(0.,){/Straight}{-1}
\FALabel(16.7452,15.5489)[b]{$b$}
\FAProp(20.,5.)(14.,6.)(0.,){/Straight}{1}
\FALabel(16.7452,4.45109)[t]{$b$}
\FAProp(6.6,10.)(14.,14.)(0.,){/Sine}{1}
\FALabel(9.85274,13.2354)[br]{$W$}
\FAProp(6.6,10.)(14.,6.)(0.,){/ScalarDash}{-1}
\FALabel(9.85274,6.76457)[tr]{$\phi$}
\FAProp(14.,14.)(14.,6.)(0.,){/Straight}{-1}
\FALabel(15.274,10.)[l]{$u_i$}
\FAVert(6.6,10.){0}
\FAVert(14.,14.){0}
\FAVert(14.,6.){0}

\FADiagram{T1 C1 N14}
\FAProp(0.,10.)(6.6,10.)(0.,){/Sine}{0}
\FALabel(3.3,8.93)[t]{$Z$}
\FAProp(20.,15.)(14.,14.)(0.,){/Straight}{-1}
\FALabel(16.7452,15.5489)[b]{$b$}
\FAProp(20.,5.)(14.,6.)(0.,){/Straight}{1}
\FALabel(16.7452,4.45109)[t]{$b$}
\FAProp(6.6,10.)(14.,14.)(0.,){/Sine}{1}
\FALabel(9.85274,13.2354)[br]{$W$}
\FAProp(6.6,10.)(14.,6.)(0.,){/Sine}{-1}
\FALabel(9.85274,6.76457)[tr]{$W$}
\FAProp(14.,14.)(14.,6.)(0.,){/Straight}{-1}
\FALabel(15.274,10.)[l]{$u_i$}
\FAVert(6.6,10.){0}
\FAVert(14.,14.){0}
\FAVert(14.,6.){0}
\end{feynartspicture}
\end{scriptsize}
\end{quote}

The fields, their propagators, and their couplings are defined in
a special file, the model file, which the user can supply or modify.
The following model files are included in \FA: the electroweak Standard
Model ({\tt SM.mod}) \cite{De93}, the same including QCD
({\tt SMQCD.mod}), and in the background-field formulation
({\tt SMbgf.mod}) \cite{DeDW95}. These model files all include the full
set of one-loop counter terms. A model file for the Minimal
Supersymmetric Standard Model (MSSM) will be available soon.

The graphics output of {\tt Paint} can be saved in a file with the
standard \mma\ functions {\tt Display} and {\tt Export}, \eg the following
two lines draw the diagrams contained in the variable {\tt ins} and save
them in a PostScript file:
\begin{quote}
\begin{verbatim}
diags = Paint[ins];
Display["diags.ps", diags]
\end{verbatim}
\end{quote}
Finally, the analytic expressions for the diagrams are obtained by
\begin{quote}
\begin{verbatim}
amp = CreateFeynAmp[ins]
\end{verbatim}
\end{quote}
The output of {\tt CreateFeynAmp} requires a detailed discussion which is
deferred to Sect.\ \ref{sect:frules}.

Needless to say, many details have been omitted in this brief survey of
\FA. All functions and their options are however fully documented in
the \FA\ manual which is included in the \FA\ package (see Sect.\
\ref{sect:download} for download information).


\section{Concepts of diagram generation}
\label{sect:concepts}

\subsection{Generation of topologies}

{\tt CreateTopologies} generates all distinct topologies for a given
number of loops and external legs. This is a purely topological process
with essentially no physics input. (If a model allows vertices of a
degree larger than four, this has to be specified explicitly.)

\begin{figure}
\begin{center}
\begin{picture}(255,150)(-30,-30)
\SetOffset(-40,-5)
\BCirc(30,50){15}

\SetOffset(-52,-5)
\Text(72,50)[l]{$\longrightarrow$}

\SetOffset(-64,-5)
\BCirc(130,50){15}
\SetWidth{1.5}
\Line(130,35)(130,10)
\SetWidth{.5}
\Vertex(130,35){2}

\SetOffset(-74,-5)
\Text(172,75)[l]{$\nearrow$}
\Text(172,50)[l]{$\longrightarrow$}
\Text(172,25)[l]{$\searrow$}

\SetOffset(-75,15)
\Line(210,80)(235,80)
\BCirc(250,80){15}
\SetWidth{1.5}
\Line(265,80)(290,80)
\SetWidth{.5}
\Vertex(235,80){2}
\Vertex(265,80){2}

\Line(210,30)(260,30)
\BCirc(275,30){15}
\Vertex(260,30){2}
\SetWidth{1.5}
\Line(210,50)(235,30)
\SetWidth{.5}
\Vertex(235,30){2}

\BCirc(250,-20){15}
\Line(210,-35)(250,-35)
\SetWidth{1.5}
\Line(250,-35)(290,-35)
\SetWidth{.5}
\Vertex(250,-35){2}
\end{picture}
\end{center}
\vskip -15bp%
\caption{\label{fig:topalg}The algorithm of {\tt CreateTopologies}, shown
for a one-loop starting topology to which two external legs are added. The
leg added in each step is drawn with a thick line.}
\end{figure}
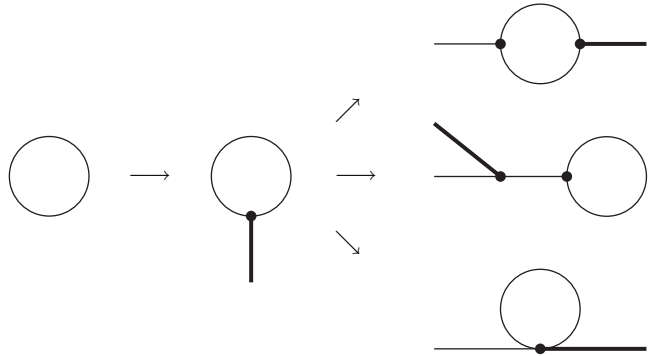

The topologies are created with a recursive algorithm \cite{KuBD90}: it
takes pre-defined topologies with zero external legs, the so-called
starting topologies, and successively adds legs until the desired number
of external legs is reached (see Fig.\ \ref{fig:topalg} for an example).
The starting topologies have to be entered once for every loop order, and
by default, \FA\ knows the tree-level, one-loop, two-loop, and three-loop
starting topologies. Note that this algorithm is not self-contained
because it depends on external input in the form of starting topologies.
The available numbers of loops are however quite sufficient for most
applications in theories like the electroweak Standard Model.

{\tt CreateTopologies} has two companion functions,
{\tt CreateCTTopologies} and {\tt Create\-VF\-Topologies}, for creating
counter-term topologies and topologies with placeholders for one-particle
irreducible vertex functions (skeleton diagrams), respectively. They use
the same algorithm as {\tt CreateTopologies}. In contrast to the ordinary
topologies, however, these topologies may contain two-vertices (\eg for
propagator-type counter terms), and the number of such vertices is not
topologically restricted by the input parameters (number of loops and
external legs). Therefore, one has to keep track of the total order of a
diagram, \ie loop order plus counter-term/vertex-function order. This is
solved in \FA\ by defining additional counter-term starting topologies.
For example, in addition to the two-loop starting topologies there exist
starting topologies with one loop and first-order counter terms, and
tree-level starting topologies with one second-order or two first-order
counter terms.

Finally, after generating all possible topologies, the topologically
equivalent ones have to be weeded out and the symmetry factors of the
remaining ones adjusted accordingly. To this end, the topologies are
sorted into some canonical order and then compared. However, this simple
procedure may fail to detect equivalent graphs if there are groups of
vertices whose permutation does not change the graph. In that case, the
indices of these symmetrical vertices (which are especially tagged in the
starting topologies for this purpose) have to be permuted to give all
topologically equivalent versions. It is this ``power set'' of each
topology that is actually compared. While this sounds like an
uncomfortably slow procedure, the actual performance is not so bad. For
example, a modern PC takes about 10 seconds to generate all 2214 two-loop
$2\to 2$ topologies.

The symmetry factor of a topology is determined in the following way: the
starting topology carries an integer $s$, the inverse of its symmetry
factor. The starting topology always has the highest $s$ because adding
legs can only diminish its symmetry. After adding each leg, the
topologically equivalent versions are gathered in the result and replaced
by one representative whose $s$ is divided by the number of versions
found.

\subsection{Field levels}
\label{sect:levels}

An important feature of \FA\ is that it distinguishes three levels of
fields:

-- {\em Generic level}, \eg the fermion {\tt F},

-- {\em Classes level}, \eg the down-type quark {\tt F[4]},

-- {\em Particles level}, \eg the $b$-quark {\tt F[4,\,\lbrac 3\rbrac]}.

\noindent
This is a quite natural concept in field theory (compare \eg the Feynman
rules in \cite{De93}) and has three enormous benefits in practical
calculations.

The kinematical structure of a coupling is fixed once the generic fields
are specified. For example, all fermion--fermion--scalar couplings are of
the form
\begin{equation}
\label{eq:ffs}
C(F,F,S) = G_-\omega_- + G_+\omega_+
\equiv\vec G\cdot\begin{pmatrix}
\omega_- \\
\omega_+
\end{pmatrix}
\end{equation}
where $\omega_\pm = (1\pm\gamma_5)/2$ are the chirality projectors. This
means that most algebraic simplifications, like the tensor reduction, need
to be carried out on the generic-level amplitude only.

The classes level saves further CPU time because the sum over a particle
index (\eg a fermion-generation index) can be performed much more
efficiently, say in a Fortran program, than the full computation of the
proportionate number of diagrams at particles level.

Thirdly, since the kinematical structure of a coupling is dictated by the
choice of representation of the Poincar\'e group (which is not often
changed), it is very profitable to store the kinematical structure apart
from the actual coupling constants so that it can be used with more than
one model. \FA\ stores the kinematical structure of the couplings in a
file called the {\em generic model file}. For example, the entry
corresponding to Eq.\ \eqref{eq:ffs} is
\begin{quote}
\begin{alltt}
AnalyticalCoupling[ s1 F[i,\,mom1], s2 F[j,\,mom2], s3 S[k,\,mom3] ] ==
  G[1][s1 F[i], s2 F[j], s3 S[k]] .
    \lbrac{} NonCommutative[ChiralityProjector[-1]],
      NonCommutative[ChiralityProjector[+1]] \rbrac{}
\end{alltt}
\end{quote}
Like Eq.\ \eqref{eq:ffs}, the right-hand side of this equation is the dot
product of an as yet unspecified vector of coupling constants $\vec G$
with the kinematical vector $(\omega_-, \omega_+)$, \ie the components of
$\vec G$ in this case provide the prefactors of $\omega_-$ and $\omega_+$,
respectively.

The actual coupling vector $\vec G$ is then specified in the {\em classes
model file}, \eg the $\overline{\ell}_{j_1}\nu_{j_2}\phi^-$-coupling in
the electroweak Standard Model is defined as
\begin{quote}
\begin{alltt}
C[ -F[2,\,\lbrac{}j1\rbrac{}], F[1,\,\lbrac{}j2\rbrac{}], S[3] ] == 
  -I EL/(Sqrt[2] SW) Mass[F[2,\,\lbrac{}j1\rbrac{}]]/MW IndexDelta[j1, j2] *
  \lbrac{} \lbrac{}1, dZe1 - dSW1/SW + dMWsq1/(2 MW^2) + dMf1[2,\,\lbrac{}j1\rbrac{}]/Mass[F[2,\,\lbrac{}j1\rbrac{}]] -
        1/2 (Conjugate[dZfR1[2,\,\lbrac{}j1\rbrac{}]] + dZfL1[1,\,\lbrac{}j1\rbrac{}])\rbrac{},
    \lbrac{}0, 0\rbrac{} \rbrac{}
\end{alltt}
\end{quote}
The outer braces on the right-hand side delimit the coupling vector, with
components corresponding to $\omega_-$ and $\omega_+$, while the inner
braces host the orders of the coupling, \eg {\tt 1} for the tree-level
coupling and ({\tt dZe1\,-\,...}) for the first-order counter term. An
overall factor ({\tt -I\,EL...}) is pulled out for clarity, but of course
multiplies all components. Incidentally, the neutrino's left-handedness
can clearly be seen from the fact that the second component of $\vec G$,
multiplying $\omega_+$, is zero in all orders. \FA\ puts almost no
restrictions on what can appear in a coupling. Indeed, most of the symbols
appearing in the example have been chosen by the model file's creator and
have no specific meaning to \FA.

Note that as yet the class indices are specified (\eg {\tt F[1]}), but
not the particle indices ({\tt\lbrac j1\rbrac}). No extra model file is
needed for particles level, however---the replacement of the remaining
particles indices by integers is trivial enough to be performed without
further input.

\subsection{Insertion of fields}

The computer-algebraic generation of Feynman diagrams corresponds to the
distribution of fields over the topologies in such a way that the
resulting diagrams possess the external fields the user has chosen and
contain only couplings allowed by the model. This process is called
``inserting fields into a topology'' and is performed by the
{\tt InsertFields} function.

As would be expected from the level-concept of fields, the insertion of
fields is a three-stage process, with functions for the insertion of
generic-, classes-, and particles-level fields nested inside each other.
It suffices however to describe the main insertion function which is
eventually invoked at all levels. This function works as follows: it is
called for each propagator and receives as input the fields coming in at
either end of the propagator, $\{f_a, f_b, \dots\}$ and $\{f_s, f_t,
\dots\}$, and the field running on the propagator itself, $f_i$:
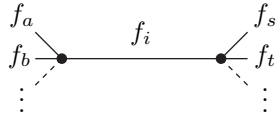
\begin{center}
\begin{picture}(100,45)(0,-5)
\Line(20,20)(80,20)
\Text(50,25)[b]{$f_i$}
\Line(20,20)(10,30)
\Text(5,32)[b]{$f_a$}
\Line(20,20)(10,20)
\Text(4,21)[]{$f_b$}
\DashLine(20,20)(10,10){2}
\Text(5,15)[t]{$\vdots$}
\Line(80,20)(90,30)
\Text(97,32)[b]{$f_s$}
\Line(80,20)(90,20)
\Text(97,21)[]{$f_t$}
\DashLine(80,20)(90,10){2}
\Text(97,15)[t]{$\vdots$}
\Vertex(20,20){2}
\Vertex(80,20){2}
\end{picture}
\end{center}
All of this field information is specified as precisely as known at
that stage of the insertion process. The insertion function then looks up
which particles are allowed for $f_i$, given that it joins $\{f_a, f_b,
\dots\}$ at the left end, and similarly for the right end. Taking the
intersection of these two ``allowed'' lists yields the possible choices 
for $f_i$.

Mixing propagators introduce a slight complication: instead of the field
$f_i$ itself, \FA\ has to take the left and right partner of $f_i$ for
look-up at the left and right end, respectively. For example, the left and
right partners of a $\gamma$--$Z$ mixing field are $\gamma$ and $Z$.

The look-up tables used for finding the allowed fields obviously play a
very important role. They are built during the model-initialization phase
and account for most of the speed of the {\tt InsertFields} function.

\subsection{Drawing Feynman diagrams}

Both the bare topologies of {\tt CreateTopologies} and the inserted
diagrams of {\tt InsertFields} can be drawn with the {\tt Paint} function.
The output of {\tt Paint}, displayed also on screen by default, can be
rendered in PostScript, \LaTeX, or any other graphics format known to
\mma, \eg GIF, JPEG, PDF, etc. The most useful output formats are however
{\tt "TeX"} (\LaTeX), {\tt "PS"} (PostScript), and {\tt "EPS"}
(encapsulated PostScript).

The \LaTeX\ format is probably the most convenient one for including the
diagrams in publications. For example, the diagram
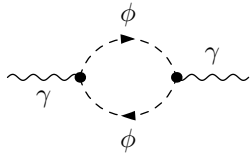
\vspace*{-5ex}
\begin{center}
\unitlength=1bp%
\begin{feynartspicture}(100,100)(1,1)
\FADiagram{}
\FAProp(0.,10.)(6.,10.)(0.,){/Sine}{0}
\FALabel(3.,8.93)[t]{$\gamma$}
\FAProp(20.,10.)(14.,10.)(0.,){/Sine}{0}
\FALabel(17.,11.07)[b]{$\gamma$}
\FAProp(6.,10.)(14.,10.)(0.8,){/ScalarDash}{-1}
\FALabel(10.,5.73)[t]{$\phi$}
\FAProp(6.,10.)(14.,10.)(-0.8,){/ScalarDash}{1}
\FALabel(10.,14.27)[b]{$\phi$}
\FAVert(6.,10.){0}
\FAVert(14.,10.){0}
\end{feynartspicture}
\end{center}
\vspace*{-5ex}
has the \LaTeX\ representation
\begin{quote}
\begin{verbatim}
\begin{feynartspicture}(100,100)(1,1)
\FADiagram{}
\FAProp(0.,10.)(6.,10.)(0.,){/Sine}{0}
\FALabel(3.,8.93)[t]{$\gamma$}
\FAProp(20.,10.)(14.,10.)(0.,){/Sine}{0}
\FALabel(17.,11.07)[b]{$\gamma$}
\FAProp(6.,10.)(14.,10.)(0.8,){/ScalarDash}{-1}
\FALabel(10.,5.73)[t]{$\phi$}
\FAProp(6.,10.)(14.,10.)(-0.8,){/ScalarDash}{1}
\FALabel(10.,14.27)[b]{$\phi$}
\FAVert(6.,10.){0}
\FAVert(14.,10.){0}
\end{feynartspicture}
\end{verbatim}
\end{quote}
Such fragments can be inserted into a \LaTeX\ document, thus eliminating
external files for the figures. It is also fairly easy to change or move
around diagrams with any text editor. The only requirement is to include
the {\tt feynarts.sty} style in which the \LaTeX\ commands used in the
\FA\ output are defined.

As of version 3, \FA\ possesses a custom PostScript prologue---a piece of
PostScript code that explains to the PostScript interpreter how to draw
propagators, vertices, and labels. The prologue makes it possible to
produce \LaTeX\ output of the form shown above and indeed,
{\tt feynarts.sty} consists mostly of the prologue. As a nice side-effect,
the PostScript files generated by \FA\ 3 are smaller by a factor of 5 or
more compared with older versions.

The shapes of the diagrams are not automatically designed by \FA. That is
a matter of human taste and too complicated for a computer program. Each
time a diagram is drawn, \FA\ looks up its shape in a database, and if no
shape is found, calls up the topology editor in which the user can arrange
the vertices, propagators, and labels with the mouse. The topology editor
is the only part of \FA\ not written in \mma\ but in Java.

\subsection{Applying the Feynman rules}
\label{sect:frules}

Once the possible combinations of fields have been determined by
{\tt InsertFields}, the Feynman rules must be applied to produce the
actual amplitudes. This is done by {\tt CreateFeynAmp}. The Feynman rules,
more specifically, consist of the expressions for the propagators and
vertices defined in the model files, a prefactor which includes symmetry
factors and usually depends on the number of loops, and the rules for the
concatenation and signs of fermion chains.

The output of {\tt CreateFeynAmp} is intentionally very symbolic to make
it easier for other programs to locate certain parts of the amplitude. For
example, the amplitude resulting from the photon self-energy diagram
painted in the last section is
\begin{quote}
\begin{tt}
\begin{flushleft}
FeynAmp[\\[.5ex]
~~GraphID[Topology~==~1,~Generic~==~1],
	\dotfill\ding{192}\\[1ex]
~~Integral[q1],
	\dotfill\ding{193}\\[1ex]
~~(\,$\dfrac{\tt I}{\tt 32~Pi^4}$~RelativeCF
	\dotfill\ding{194}\\
~~~~FeynAmpDenominator[$\dfrac{\text{1}}{\tt
q1^2\text{~-~}\text{Mass[S[Gen3]]}^2}$,\\
~~~~~~$\dfrac{\text{1}}{\tt
\text{(-p1\,+\,q1)}^2\text{~-~}\text{Mass[S[Gen4]]}^2}$]
	\dotfill\ding{195}\\[1ex]
~~~~(p1~-~2~q1)[Lor1]~(-p1~+~2~q1)[Lor2]
	\dotfill\ding{196}\\[1ex]
~~~~ep[V[1],\,p1,\,Lor1]~ep$^{\text{*}}$[V[1],\,k1,\,Lor2]
	\dotfill\ding{197}\\[1ex]
~~~~G$\tt ^{\text{(0)}}_{SSV}$[(Mom[1]\,-\,Mom[2])[KI1[3]]]
	\dotfill\ding{198}\\
~~~~G$\tt ^{\text{(0)}}_{SSV}$[(Mom[1]\,-\,Mom[2])[KI1[3]]]\,),\\[1ex]
~~\lbrac\,Mass[S[Gen3]],~Mass[S[Gen4]],
	\dotfill\ding{199}\\
~~~~G$^{\text{(0)}}_{\text{SSV}}$[(Mom[1]\,-\,Mom[2])[KI1[3]]],\\
~~~~G$^{\text{(0)}}_{\text{SSV}}$[(Mom[1]\,-\,Mom[2])[KI1[3]]],\\
~~~~RelativeCF\,\rbrac~->\\
~~Insertions[Classes][\lbrac MW,~MW,~I~EL,~-I~EL,~2\rbrac]\\
]
\end{flushleft}
\end{tt}
\end{quote}
The {\tt FeynAmp} function has four arguments: an identifier \ding{192},
the integration momenta \ding{193}, the generic amplitude
\ding{194}--\ding{198}, and replacement rules for transforming the generic
amplitude into a classes or particles amplitude \ding{199}.

The generic amplitude has the following elements: a numeric pre-factor
\ding{194} ({\tt RelativeCF} stands for {\sc relative} {\sc c}ombinatorial
{\sc f}actor and is specified for each diagram by the replacement rules
\ding{199}), the denominators of loop propagators collected in a function
{\tt FeynAmpDenominator} \ding{195}, the kinematic structure of the two
scalar--scalar--vector couplings \ding{196}, the polarization vectors of
the external photons \ding{197}, and the generic coupling constants of
both vertices \ding{198}.

To turn the generic amplitude into a classes or particles amplitude, all
generic objects must be replaced by their concrete values at the
particular level. This replacement is specified by the rules \ding{199}.
For example, for the first (and in this simple example only) classes
diagram, {\tt Mass[S[Gen3]]} becomes {\tt MW}. In general, one generic
amplitude will of course fan out into several derived classes or particles
amplitudes, so the {\tt Insertions} function will have several entries.

Even though this method of keeping the generic amplitude apart from the
replacement rules has the advantages outlined in Sect.\ \ref{sect:levels},
it is possible to obtain the more conventional Feynman amplitudes by
picking out one level (\ie applying the replacement rules) with the
function {\tt PickLevel}.

From the appearance of polarization vectors it is clear that the sample 
amplitude shown above is part of an $S$-matrix element. Just as well it is
possible to produce amplitudes for Green's functions by selecting
{\tt Truncated -> True} as an option in {\tt CreateFeynAmp}.

\subsection{Supersymmetric models}

Supersymmetric theories in general contain Majorana fermions and hence
fermion-number-violating couplings (\eg quark--squark--gluino). The
textbook prescription of ordering the Dirac matrices opposite to their
occurrence along the arrows on fermionic lines obviously breaks down in
this case since one cannot define a fermion-number flow. Put differently,
Majorana-fermion lines have no arrow.

In fact, \FA\ has to address this problem for all fermions, not just for
Majorana ones, because the amplitude is constructed at generic level and
generic fermion fields are defined to be undirected.

The implemented solution is the ``flipping-rule'' algorithm
\cite{DeEHK92}: instead of traversing the fermion lines along the
fermion-number flow imposed from the outside, \FA\ chooses a direction for
each fermion chain. If it turns out later that, for a Dirac fermion, the
chosen direction is opposite to the actual fermion flow, it ``flips'' the
coupling, \ie it derives the coupling appropriate for the reversed
fermion flow from the known coupling. This is in fact nothing but a
charge (as opposed to hermitian) conjugation of the coupling and the
flipping rules, which act on elements of the Dirac algebra, are actually 
quite simple, \eg
\begin{equation}
\gamma_\mu\omega_\pm
\overset{\text{flip}}{\longrightarrow}
-\gamma_\mu\omega_\mp\,.
\end{equation}


\section{Availability, Requirements}
\label{sect:download}

The \FA\ package can be downloaded from {\tt http://www.feynarts.de}
and includes a comprehensive manual which explains installation and
usage. \FA\ requires \mma\ 3 or above. For the topology editor, a Java
VM and the J/Link package are needed, both of which can be obtained free
of charge (see the instructions on the web site). \FA\ is an open-source
program and stands under the GNU library general public license.

\section*{Acknowledgements}

The development of \FA\ 2.2 has been supported by the Deutsche
Forschungsgemeinschaft (Forschergruppe ``Quantenfeldtheorie,
Computeralgebra und Monte-Carlo Simulation'') under contract number Ku
502/8--1. The new graphics functions of \FA\ 3 have been developed on a
Visiting Scholar grant of Wolfram Research, Inc., and I am grateful in
particular to M.\ Malak and L.\ D'Andria for discussions on how to
get the most out of \mma. I thank A.\ Denner and W.\ Hollik for
proofreading the manuscript.


\newcommand{\cpc}[3]{{\sl Comp. Phys. Commun.} {\bf #1} (#2) #3}
\newcommand{\fp}[3]{{\sl Fortschr. Phys.} {\bf #1} (#2) #3}
\newcommand{\np}[3]{{\sl Nucl. Phys.} {\bf #1} (#2) #3}
\newcommand{\pl}[3]{{\sl Phys. Lett.} {\bf #1} (#2) #3}
\newcommand{\pr}[3]{{\sl Phys. Rev.} {\bf #1} (#2) #3}
\newcommand{\zp}[3]{{\sl Z. Phys.} {\bf #1} (#2) #3}
\newcommand{\nim}[3]{{\sl Nucl. Instr. Meth.} {\bf #1} (#2) #3}
\newcommand{\jcp}[3]{{\sl J. Comput. Phys.} {\bf #1} (#2) #3}
\newcommand{\ptps}[3]{{\sl Prog. Theor. Phys. Suppl.} {\bf #1} (#2) #3}

\end{document}